\documentclass[runningheads]{llncs}
\usepackage[misc,geometry]{ifsym}
\usepackage[T1]{fontenc}
\usepackage{xcolor}
\usepackage{graphicx}
\usepackage{multirow}
\usepackage{enumitem}
\usepackage{subfigure} 
\usepackage{url}

\begin{document}
\title{A Federated Learning Framework for Stenosis Detection}

\author{Mariachiara Di Cosmo\inst{1}$\spadesuit$ 
\and Giovanna Migliorelli\inst{2}
\and Matteo Francioni\inst{3}
\and Andi Mu\c{c}aj\inst{3}
\and Alessandro Maolo\inst{3}
\and Alessandro Aprile\inst{4}
\and Emanuele Frontoni \inst{5}
\and Maria Chiara Fiorentino\inst{1}
\and Sara Moccia\inst{6}
}

\authorrunning{M. Di Cosmo et al.}

\institute{Department of Information Engineering, Università Politecnica delle Marche, Ancona, Italy 
\and Department of Law, Università degli Studi di Macerata, Macerata, Italy  
\and U.O.C. Cardiology and Hemodynamics - Department of Cardiovascular Sciences, Azienda Ospedaliero Universitaria delle Marche, Ancona, Italy
\and U.O.C. Cardiology-UTIC-Hemodynamics - Ospedale del Mare, ASL NA 1, Napoli
\and Department of Political Sciences, Communication and International Relations, University of Macerata, Macerata, Italy
\and The BioRobotics Institute and Department of Excellence in Robotics and AI, Scuola Superiore Sant'Anna, Pisa, Italy
\\ Corresponding author: $\spadesuit$ \email{m.dicosmo@pm.univpm.it}}
\maketitle

\begin{abstract}

This study explores the use of Federated Learning (FL) for stenosis detection in coronary angiography images (CA).
Two heterogeneous datasets from two institutions were considered: Dataset 1 includes 1219 images from 200 patients, which we acquired at the Ospedale Riuniti of Ancona (Italy); Dataset 2 includes 7492 sequential images from 90 patients from a previous study available in the literature. Stenosis detection was performed by using a Faster R-CNN model. In our FL framework, only the weights of the model backbone were shared among the two client institutions, using Federated Averaging (FedAvg) for weight aggregation. 
We assessed the performance of stenosis detection using Precision ($Prec$), Recall ($Rec$), and F1 score ($F1$).
Our results showed that the FL framework does not substantially affects clients 2 performance, which already achieved good performance with local training; 
for client 1, instead, FL framework increases the performance with respect to local model of +3.76\%, +17.21\% and +10.80\%, respectively, reaching  $Prec=73.56$, $Rec=67.01$ and $F1=70.13$. 
With such results, we showed that FL may enable multicentric studies relevant to automatic stenosis detection in CA by addressing data heterogeneity from various institutions, while preserving patient privacy.

\end{abstract}

\section{Introduction}
\label{sec:Introduction}

Coronary artery disease (CAD) provokes stenoses, coronary segments with narrowed lumen, causing blood flow reduction and eventually leading to ischemia and heart attacks. CAD, which is a leading cause of mortality worldwide, is currently assessed through coronary angiography (CA), an imaging technique that uses X-rays and contrast dye to visualize coronary arteries and assess blood flow dynamics \cite{saraste2020esc}. Precise and timely stenosis detection is crucial for effective CAD diagnosis and treatment. CA interpretation relies on clinician's expertise and requires tackling challenges such as complex vessel anatomy, stenosis variability (in shape, pattern and severity), presence of movement and shadowing artifacts, as well as varying imaging equipment and contrast agent levels \cite{lawton20222021}.

Computer-aided decision support systems for stenosis detection from CA have the potential to improve CAD assessment and reduce clinician variability.
In recent years, deep learning (DL) has shown great potential in automating stenosis detection from CA images. 
Many approaches propose a multi-stage framework, in which stenoses are identified following vessel enhancement \cite{moon2021automatic} or segmentation \cite{zhao2021automatic}. 
In \cite{zhao2021automatic}, a U-Net++ model with a feature pyramid network is proposed to automatically segment coronary arteries and from the artery centerline, diameters are calculated and stenotic levels are measured. 
In \cite{moon2021automatic}, after key frame detection using vessel extraction, a classification model identifies the stenosis presence in the key frame and through Class Activation Mapping (CAM) stenosis is qualitatively localized. The study considers right CA only. 

To avoid error accumulation though image pre-processing steps, several studies develop DL methods to localize stenoses directly from CA images without relying on vessel analysis \cite{cong2019automated,danilov2021real,ovalle2022hybrid}. 
In \cite{cong2019automated}, similarly to \cite{moon2021automatic}, CAM is employed to localize stenosis on top of an image-level stenosis classification performed by an Inception-v3.  
Several object detection models are trained in \cite{danilov2021real} and tested over one-vessel stenotic CA series to explore trade off between accuracy and performance efficiency. 
Inspired by quantum computing, \cite{ovalle2022hybrid} incorporates a quantum network in a ResNet network to detect stenosis by performing binary classification on fixed-size patches obtained from CA images.
Some approaches \cite{zhang2019direct,wu2020automatic,han2023coronary,pang2021stenosis} consider sequence of images to take advantage of temporal information intrinsic in CA. 
The work in \cite{zhang2019direct} proposes a hierarchical attentive multi-view learning model to capture the pixel correlation from CA sequences for quantifying stenosis. 
In \cite{wu2020automatic}, a method is proposed that detects candidate stenoses using a deconvolutional single-shot detector (DSSD) on frames of a CA sequence selected by a U-Net model. Then, the seq-fps module takes advantage of the temporal information within the X-ray sequence to suppress false positives and generate the final stenosis detection results. 
The work in \cite{pang2021stenosis} includes sequence feature fusion and sequence consistency alignment into stenosis detection network to capture the spatio-temporal information using 166 CA sequence. 
To gather spatio-temporal features, a transformer-based module is used in \cite{han2023coronary} and to learn long-range context a feature aggregation network is developed.
Most of these studies use proprietary datasets which have their own acquisition and annotation processes, or open-source datasets which are made available for research community (for stenosis detection only \cite{danilov2021real} shares dataset). 
This limits generalizability as DL model results relies on the dataset-specific characteristics such as imaging equipment, protocols, and patient variability.

When multicentric studies are considered \cite{cong2019automated,pang2021stenosis,zhang2019direct}, datasets are treated separately, simply testing their model on each dataset overlooking the potential influence of one dataset over the other, or as a single dataset, without considering issues related to sharing medical imaging data across multiple centers.
Data privacy concerns, regulatory compliance, technical compatibility, and varying imaging protocols are addressed among the main challenges to face for DL-based decision support tool development for medical image analysis \cite{fiorentino2022review,nazir2023federated}.
Federated learning (FL) \cite{sheller2020federated} provides an opportunity for data-private multi-institutional collaborations, leveraging model learning sharing among datasets interactions while preserving their privacy. 
Each client’s raw data are stored locally and remain under control of and private to that institution, while only model updates leave the client enabling the aggregation of learned patterns into a single global model. In addition, different FL strategies can be adopted to increase data privacy further, like partial model sharing \cite{yang2021flop}, to proactively avoid data leakage during communication between the server and multiple clients. 
FL has already shown to be successful in medical image analysis \cite{xiao2022catenorm,lu2022federated}: it has been used to alleviate small sample size and lack of annotation problems, facilitate domain adaptation and mitigate domain shift. Despite its potential, no applications in stenoses detection have been proposed so far. 

The present work explores FL for stenosis detection from two datasets with the aim of addressing challenges posed by data heterogeneity, variable acquisition protocols and inter-patient variability. 
Our contributions include:
\begin{itemize}
    \item FL framework to address stenosis detection from CA;
    \item Privacy-preserving FL by partial aggregation of the detection model to capture and share intrinsic CA features;
    \item Collection of a new dataset of CA, composed by 1219 images from 200 patients and acquired in the clinical practice from three clinicians, who performed image annotation.
\end{itemize}

\section{Material and Methods}
\label{sec:Methods}

\begin{figure}
\centering
\includegraphics[width=0.8\textwidth]{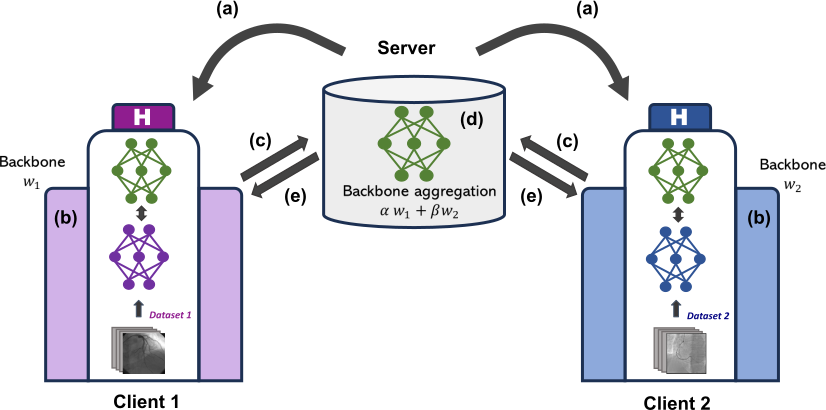}
\caption{Representation of proposed Federated Learning (FL) framework: (a) the server initially sends the Faster R-CNN model pre-trained on Coco dataset to both clients; (b) for each round, the model is trained locally on the training set of the specific client; (c) updated weights of the model backbone are sent back to the server; (d) the server receives and aggregates the updated weights from both clients; (e) the server sends the result of aggregation back to the two clients.
Communication happens between server and clients aggregating at each round the backbone weights ($w_1$ for client 1 and $w_2$ for client 2) with Federate Averaging (FedAvg) defined as $\alpha w_1 + \beta w_2$, where $\alpha$ and $\beta$ represent two parameters that determine the clients contribution to the aggregation process. } \label{fig1}
\end{figure}

Our FL framework relies on a learning paradigm between two hospitals (i.e., the local clients of the federation) and one central server. 

To perform stenosis detection a Faster R-CNN model \cite{Ren2015FasterRT} is deployed. Faster R-CNN integrates a Convolutional Neural Network (CNN) as backbone for feature extraction, a Region Proposal Network (RPN) for proposal generation, and subsequent layers for accurate object classification and localization within proposed Region of Interest (ROI).
In the present study, the model developed is a Faster R-CNN \cite{Ren2015FasterRT} with ResNet50 Feature Pyramid Network (FPN) as backbone network \cite{li2021benchmarking}. This version of Faster R-CNN benefits from the hierarchical features extracted by the ResNet50, the multi-scale feature representation provided by FPN, an extension of the RPN with two convolutional layers on top of the ROI heads for classification and regression of the stenotic regions.
The Faster R-CNN, pre-trained on Coco dataset \cite{lin2014microsoft}, is trained to minimize a multi-task loss as a weighted combination of cross-entropy loss for the identification of stenosis presence and regression loss for stenosis localization.
Since the backbone is responsible for feature extraction and representation, we consider that sharing its parameters in the FL framework could allow the clients to collectively learn generalized and meaningful features of the CA images. 

As shown in Fig. \ref{fig1}, the server initiates the federated process sending the Faster R-CNN model to both the clients, including the pre-trained backbone weights. Then, for each round of the federated computation, each client trains the model received on its local private data and sends back to the server only the updated backbone weights. The server receives the weight updates from each client and aggregates them to create a global model. The aggregation technique adopted relies on Federated Averaging (FedAvg) \cite{mcmahan2017communication}, which computes a weighted average of the backbone parameters collected by the server. Since only the backbone weights are shared, the server combines them without interacting with the rest of the Faster R-CNN model. 

FedAvg aggregation strategy is defined as $\alpha w_1 + \beta w_2$, where $\alpha$ and $\beta$ represent two parameters that determine the clients' contribution to the aggregation process and $w_1$ and $w_2$ are the backbone weights of client 1 and client 2, respectively.

\subsection{Datasets} \label{subsec:Dataset}

\begin{figure}[tbp]
\centering
\includegraphics[width=.6\textwidth]{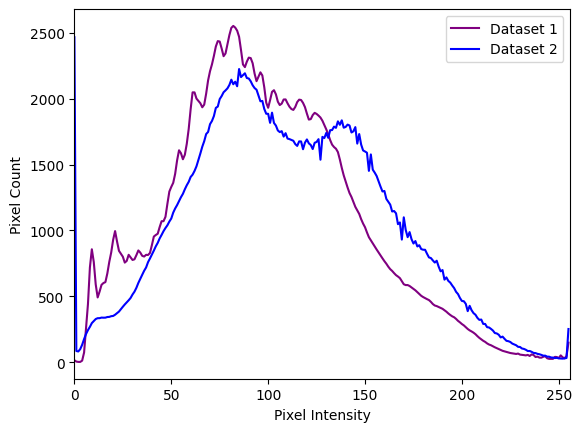}
\caption{Domain shift between the two coronary angiography (CA) datasets in terms of pixel value distribution. 
} \label{fig2}
\end{figure}

The datasets used in this study exhibit inherent dissimilarities in the acquisition protocols and annotation approaches. This results in a visible data heterogeneity: as depicted in Fig. \ref{fig2}, the intensity distributions at the two institutions underline variation in imaging characteristics posing domain shifts challenges, thus leading to reduced model generalization, increased bias and ineffective transfer learning. Domain shift arises especially from image characteristics, such as imaging equipment, image quality, frame selection process, and patient demographics.

Both datasets are made of gray-scale images with 512x512 pixels and all images acquired from the same patient are carefully considered as part of the same set (training or testing).

\subsubsection{Dataset 1}  \label{subsec:Dataset1}

The dataset consists of 1219 CA images, which we acquired at the Ospedali Riuniti of Ancona (Italy). 
The images are provided by 200 patients, who signed informed consent, underwent CA procedure and presented one or more stenotic regions along the coronary arteries (up to 4 stenoses per image). 
From each patient exam, a few relevant frames are selected by the clinicians of this study taking into account the presence of high contrast dye, various viewpoints, and potentially the diastolic phase. 
The data acquisition is conducted in compliance with the Helsinki Declaration and under the supervision of three expert clinicians, which evaluate the presence of the stenotic regions and provide the annotations.
For model training and testing, 1106 (90\%) images from 175 patients compose the training set and 112 (10\%) images from 25 patients the test set.

\subsubsection{Dataset 2}  \label{subsec:Dataset2}

The dataset is provided by Danilov et al. \cite{danilov2021real} and contains CA series comprising a total of 7492 images. The images are obtained from 100 patients, who underwent CA 
at the Research Institute for Complex Problems of Cardiovascular Diseases (Kemerovo, Russia).
All patients had confirmed one-vessel CAD.
From each patient exam, images containing contrast passage through a stenotic vessel are extracted in sequences, discarding non-informative images. 
The manual annotation of the presence or absence of stenotic lesions for each image was performed by a single operator, as described in \cite{danilov2021real}.
For model training and testing, 6660 (90\%) images from 80 patients are used as training set, and 832 (10\%) images from 10 patients as test set. The split was given by \cite{danilov2021real}.

\subsection{Experimental protocol}
\label{subsec:Setup}

In Faster R-CNN training and for all experiments, we used Adam optimizer with a constant learning rate set to 0.0001 and batch size equal to 16. Pixel values in the input images were normalized in the range [0,1], and the images were randomly augmented via horizontal flip.
To handle the different sizes of Dataset 1 and Dataset 2 and ensure comparable number of training steps, the FL framework performed 20 rounds. In each round, the training process of the Faster R-CNN was performed locally for 20 epochs for Dataset 1 and 4 epochs for Dataset 2. For the first round, a warm-up strategy was implemented to promote weight stabilization and the local training was performed more extensively for 40 and 16 epochs for Dataset 1 and Dataset 2, respectively. 
FedAvg aggregation is performed by assigning weights proportionally to the dataset sizes, defining $\alpha = 1$ and $\beta = 6$. Even though our clients strongly differ in terms of number of images, we consider that giving more importance to the larger dataset could be beneficial for the smaller one.

We compared the performance of proposed FL framework with that obtained by training the Faster R-CNN model locally. 
The local models were trained for 200 epochs for client 1 and 50 epochs for client 2.
We further performed the following ablation study:  

\begin{itemize}[leftmargin=.35in]
    \item[\textbf{FL1:}] \textit{Faster R-CNN weight aggregation}
    
    To probe the effectiveness of sharing only the weights of the backbone, focusing on extracting and sharing relevant features, we performed also the FL of the whole Faster R-CNN model to evaluate the impact on stenosis detection performance.

    \item [\textbf{FL2:}] \textit{Faster R-CNN backbone weight aggregation with equal client contribution}

    Considering the strong difference in terms of size between the two datasets, we explored also the effect of applying weighting techniques to the aggregation process to ensure a fair and balanced representation of both clients ($\alpha=1$ and $\beta=1$). In this way, we examine whether the discrepancy in terms of size between the two datasets may lead to any performance penalty in detriment of smaller datasets. 
    
    \item [\textbf{FL3:}] \textit{Faster R-CNN backbone weight aggregation with the exclusion of Batch Normalization layers parameters}

    Based on the study of \cite{li2021fedbn}, which demonstrated that keeping local Batch Normalization parameters not synchronized with the global model reduces feature shifts in non-Independent and Identically Distributed (IID) data as in our case (see Fig. \ref{fig2}), we evaluated if the exclusion of the statistical non-trainable parameters of the Batch Normalization layers of the backbone could mitigate the discrepancy between the clients.
\end{itemize}

To assess the performance of stenosis detection, we computed Precision ($Prec$), Recall ($Rec$) and F1 score ($F1$) over each client test set. We considered a prediction as True Positive ($TP$) if it achieved an Intersection over Union ($IoU$) value with respect to the ground truth annotation greater than or equal to 0.5. Conversely, a predicted bounding box with a $IoU$ value less than 0.5 was considered a False Positive ($FP$). When a stenosis did not have any corresponding prediction, it is regarded as a False Negative ($FN$). 
To ensure the lowest number of missing predictions, for each FL framework training and for each model training best configuration was selected in terms of average $Rec$ over the client' test set.

The overall implementation was performed in Python 3.8.10 with PyTorch v.2.0.0, Torchvision v.0.15.1 and Flower v.2.0.0 libraries. Our model training and all experiments conducted were performed via 8 GPU bank where one or more GPUs were assigned to each client and one GPU was assigned to the server. 

\section{Results and Discussion}
\label{sec:Results}

\begin{table}[tbp] 
\centering\label{tab1}
\caption{Mean values of Precision ($Prec$), Recall ($Rec$) and F1 score ($F1$): from top to bottom single clients' local model, \textbf{FL1}, \textbf{FL2}, \textbf{FL3} and proposed FL framework performances are reported. }
\begin{tabular}{c|c c c c|c c c c}
\multicolumn{1}{c}{}&\multicolumn{4}{c|}{\textbf{Dataset 1}}&\multicolumn{4}{|c}{\textbf{Dataset 2}}\\
\cline{2-9}
\multicolumn{1}{c}{}& $Prec$ & $Rec$ & $F1$ & \textit{Round} &  $Prec$ & $Rec$ & $F1$ & {Round}\\
\hline
\textbf{Local}&70.89&57.14&63.28& -  &   \textbf{93.50}&82.83&87.84&-\\
\textbf{FL1} &76.25&62.24&68.54& 2    &   92.61&82.71&87.38&2\\
\textbf{FL2} & \textbf{77.33}&59.18&67.05& 3    &   92.54&83.43&87.75&1\\
\textbf{FL3} &69.32&62.24&65.59& 1    &   92.91&83.43&\textbf{87.92}&5\\
\textbf{Proposed} &73.56&\textbf{67.01}&\textbf{70.13}& 2   &   91.62&\textbf{84.03}&87.66&1\\
\end{tabular}
\end{table}

$Prec$, $Rec$ and $F1$ values achieved from proposed FL framework are reported in Table \ref{tab1} in comparison with performance obtained from single client's local training and from \textbf{FL1}, \textbf{FL2} and \textbf{FL3}. Table \ref{tab1} shows that client 1 performance improved significantly within the FL framework: the $Prec$, $Rec$ and $F1$ values increased considerably compared to local training of +3.76\%, +17.21\% and +10.80\% respectively, demonstrating the positive effect of the interaction with client 2.
The sharing of the backbone weights only is noteworthy for client 1: a higher $Rec$ value compared to \textbf{FL1} (+7.66\%) is achieved, suggesting a reduction of $FN$ predictions.
Client 2 had a greater contribution in the backbone aggregation process and this introduced additional insights and increased the extraction of intrinsic features which boosted the ability of client 1 to detect stenotic regions.
In Table \ref{tab1}, it is also evident that client 2 performance remained relatively stable and unaffected within the FL process: it consistently achieved comparable results across \textbf{FL1}, \textbf{FL2}, \textbf{FL3} and proposed FL framework. The $Prec$ value was highest when the model was trained locally, indicating that the introduction of a FL framework may not significantly improve performance for client 2.
For what concerns \textbf{FL3}, differently from \cite{li2021fedbn}, client 1 did not benefit from the exclusion from the aggregation process of the Batch Normalization layers parameters, while client 2 exhibited good performance under this setting. 
With \textbf{FL3} experiment, we focused on investigating the specific impact of Batch Normalization exclusion as addressed in \cite{li2021fedbn}; however,  we are aware that in recent research \cite{wang2023batch} alternatives to Batch Normalization have been explored to mitigate challenges associated with non-IID data distribution in FL scenarios. In future work, we plan to expand our study exploring also alternative normalization techniques.

\begin{figure} [tbp]
\centering
\includegraphics[width=0.8\textwidth]{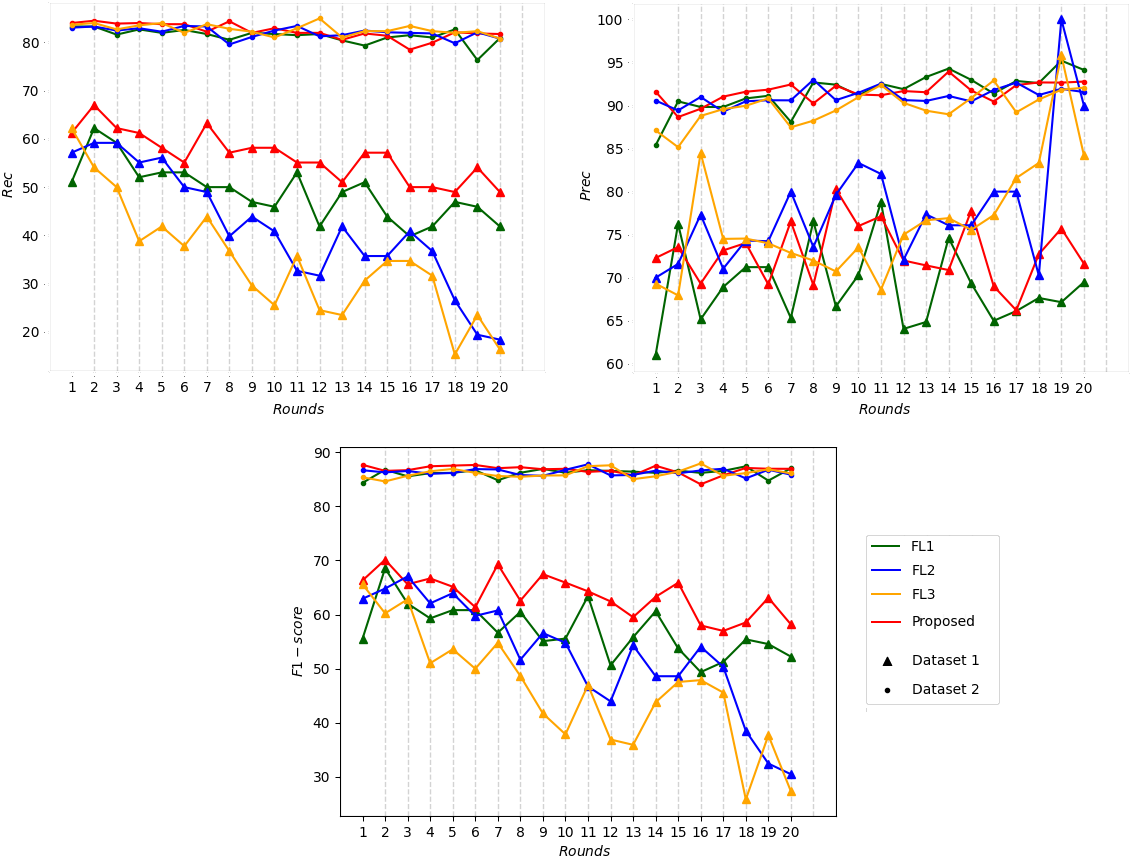}
\caption{Mean values of Precision ($Prec$), Recall ($Rec$) and F1 score ($F1$) at each round of \textbf{FL1}, \textbf{FL2}, \textbf{FL3} and proposed FL framework. } \label{fig3}
\end{figure}

Figure \ref{fig3} shows the trends of $Prec$, $Rec$ and $F1$ obtained from the proposed FL framework and from \textbf{FL1}, \textbf{FL2} and \textbf{FL3} considering different number of rounds.
The proposed FL framework confirmed to enhance client 1 performance. 
On the other hand, client 2 demonstrated remarkable results throughout the different rounds with minimal variation, suggesting that FL framework did not significantly impact over its performance.
By increasing the number of rounds, the proposed FL framework and also \textbf{FL1}, \textbf{FL2} and \textbf{FL3} exhibited sings of overfitting. It is evident from the declining performance of client 1, especially in terms of $Rec$ and $F1$. The $Prec$, instead, could have not manifested a similar decline, as it is less susceptible to overfitting compared to the other metrics. Client 2 performance, then, showed stable trends, without improvements at increasing number of rounds.
In addition to current ablation studies (\textbf{FL1}, \textbf{FL2}, \textbf{FL3}), we consider also to explore in the future the impact of uniform training epochs across communication rounds for both clients,  to evaluate the effect of FedAvg aggregation in standard settings and its influence on convergence behaviour.

\begin{figure}[tbp]
\centering
\includegraphics[width=1\textwidth]{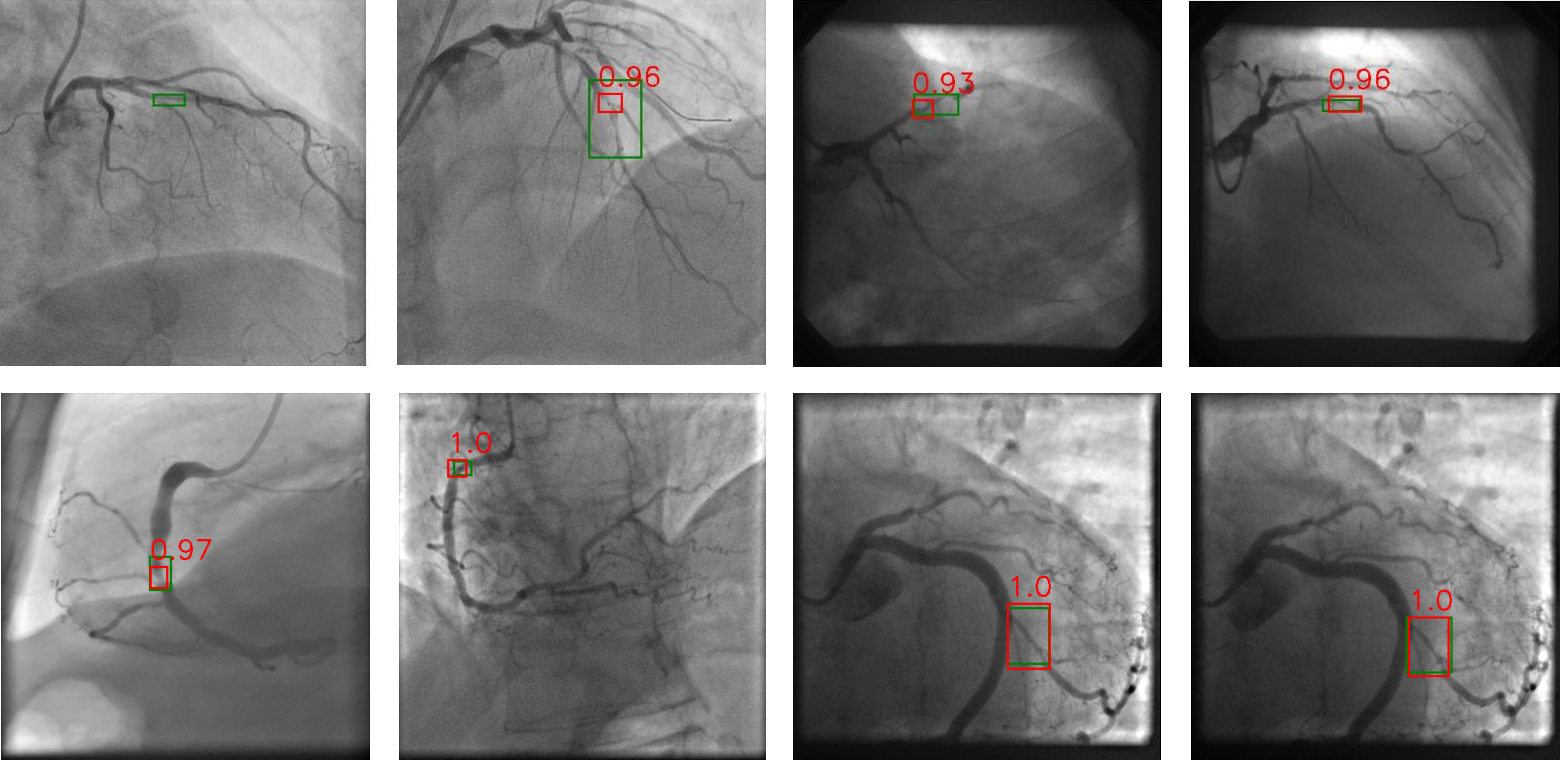}
\caption{Samples of stenosis predictions from Dataset 1 (first row) and Dataset 2 (second row) obtained with the proposed FL framework: ground truth is shown in green box, prediction with confidence score in red.} \label{fig4}
\end{figure}

Figure \ref{fig4} shows clients performance highlighting the difference between their datasets. Client 1 (first row), with its smaller and highly variable dataset, exhibited poorer performance, as visible from the presence of several missing predictions and less precise bounding box overlapping. In contrast, for client 2 (second row) stenoses are almost always recognized and localized accurately in the images. In Fig. \ref{fig4} also two images belonging to the same patient, one from Dataset 1 and one from Dataset 2, are displayed as third and forth images of the row: for each patient CA procedure, during the dataset annotation process, at client 1 only significant frames were selected, whereas at client 2 sequential frames were all annotated, disregarding only the frames in which no stenosis was visible. This difference further accentuated datasets differences and the lower variability of Dataset 2 compared to Dataset 1.  

The overall results showed the effectiveness of proposed FL framework in improving stenosis detection performance of client 1, by leveraging information provided by client 2. Sharing backbone weights allowed us to transfer knowledge from dataset 2, overcoming limits given by a small dataset size.
On the other hand, client 2, already performing well locally, was not substantially impacted by the FL framework. However, sharing information with Dataset 1, with its smaller but highly variable nature, could have been helpful for client 2 in making its model more robust and more generalizable.
In addition, the adoption of a partial model sharing approach though the aggregation of backbone weights only, as in \cite{yang2021flop}, enhanced further data privacy protection, extracting only intrinsic and general features from CA images.


FL offers several potential advantages for stenosis detection in CA. First, it enables collaborative learning across multiple institutions, allowing the inclusion of diverse datasets and facilitating the development of more generalized models \cite{sheller2020federated}. This is particularly crucial in the context of stenosis detection, as datasets can be very heterogeneous and exhibit wide variations in terms of imaging protocols, annotation procedures, patient populations, and equipment used.
By involving multiple institutions and datasets, multicentric studies could provide a more comprehensive and representative view of clinical practice.  
Moreover, FL offers privacy-preserving capabilities: by performing training process locally on individual clients data, FL mitigates the need for data sharing while still enabling collaboration and knowledge sharing among different institutions \cite{sheller2020federated}. 
This is a critical aspect in medical imaging research, where strict privacy regulations, ownership, regulatory compliance, and ethical considerations come into play \cite{myrzashova2023blockchain}. 
The significance of multicentric studies and privacy-preserving approaches is further emphasized by the growing interest and attention from organizations such as the European Commission.
A document to promote ethical principles in DL model design and deployment, called Ethics Guidelines for Trustworthy AI \footnote{\url{https://ec.europa.eu/futurium/en/ai-alliance-consultation.1.html}}, was published in 2018 from the European Commission, which also encourages the use of FL in DL development providing useful information about FL impact on data protection \footnote{\url{https://edps.europa.eu/press-publications/publications/techsonar/federated-learning_en}}.

Overall, our study highlights the importance of considering data heterogeneity and privacy concerns in the development of stenosis detection models and even though further research is needed to optimize the FL process and include multiple institutions for a wider representation of data heterogeneity, it opens the way for an efficient clinicians support to  stenosis detection form CA, ultimately leading to improved patients clinical outcomes.

\section{Conclusion}
\label{sec:Conclusion}

In this study, we explored the use of FL for stenosis detection in CA. Training with data from different institutions is particularly relevant in this context, where datasets exhibit wide variations in imaging protocols, annotation procedures, patient populations, and equipment used, in addition to the intrinsic CA imaging challenges. 
Our FL framework, by sharing the Faster R-CNN backbone weights, improved stenosis detection accuracy for client 1 achieving an increase in $Prec$, $Rec$ and $F1$ of +3.76\%, +17.21\% and +10.80\% respectively, while client 2, which already achieved high stenosis detection ability training the model locally, did not benefit significantly from a FL framework.
We hope our study may pave the way for future studies on privacy-preserving computer-assisted algorithms for CAD diagnosis.

\bibliographystyle{splncs04}
\bibliography{mybibliography}

\end{document}